 \documentclass[aps,prb,reprint,superscriptaddress]{revtex4-1}

\usepackage{amsmath,amssymb,bm}
\usepackage{graphicx}
\usepackage{hyperref}
\usepackage{float}
\usepackage{braket}
\usepackage[usenames,dvipsnames,svgnames,table]{xcolor}

\usepackage{comment}

\renewcommand\Re{{\rm Re}}
\renewcommand\Im{{\rm Im}}

\begin{document}

\title{Robustness of equilibrium off-diagonal current fluctuation against localization of electron states in macroscopic two-dimensional systems}
\author{Kentaro Kubo}
\email{kubo@as.c.u-tokyo.ac.jp}
\affiliation{Komaba Institute for Science, The University of Tokyo, 3-8-1 Komaba, Meguro, Tokyo 153-8902, Japan}
\affiliation{Department of Basic Science, The University of Tokyo, 3-8-1 Komaba, Meguro, Tokyo 153-8902, Japan}
\author{Kenichi Asano}
\email{asano@celas.osaka-u.ac.jp}
\affiliation{Center for Education in Liberal Arts and Sciences, Osaka University, Toyonaka, Osaka 560-0043, Japan}
\author{Akira Shimizu}
\email{shmz@as.c.u-tokyo.ac.jp}
\affiliation{Komaba Institute for Science, The University of Tokyo, 3-8-1 Komaba, Meguro, Tokyo 153-8902, Japan}
\affiliation{Department of Basic Science, The University of Tokyo, 3-8-1 Komaba, Meguro, Tokyo 153-8902, Japan}

\date{\today} 

\begin{abstract}
We study the off-diagonal current fluctuation in a macroscopic quantum system measured in an ideal manner that is as close as possible to the classical ideal measurement. 
We show rigorously that not only extended but also localized states contribute to the off-diagonal current fluctuation.
This result contrasts with the fact that only the extended states affect the off-diagonal (Hall) conductivity and apparently contradicts the naive expectation from the fluctuation-dissipation theorem that might directly connect these two quantities.
More specifically, we study the off-diagonal current fluctuation in a disordered two-dimensional electron system in a strong magnetic field at low temperatures.
The fluctuation is almost unchanged from that of the pure system reflecting the property mentioned above, being approximately proportional to the Landau level filling factor with high accuracy.
Our finding paves the way to estimate the filling factor and the electron density from the off-diagonal current fluctuation observed in macroscopic systems. 	
\end{abstract}

\maketitle

\section{Introduction}
\label{sec:intro}
The equilibrium fluctuation of electric current in a macroscopic system has been extensively studied from the viewpoint of the fluctuation-dissipation theorem (FDT)\cite{Einstein, Johnson, Nyquist, Onsager1,Onsager2,Takahashi,CallenWelton,Nakano,Kubo1,Kubo2,KTH}. 
According to the FDT, the equilibrium current fluctuation should agree with the product of the temperature $T=1/\beta$ (we take $k_{\rm B}=1$) and the electric conductivity in response to an infinitesimal electric field. 

While the FDT was originally proposed on the diagonal conductivity and the diagonal current fluctuation\cite{Johnson,Nyquist}, it was later proved rigorously in a classical system that the FDT also holds for the off-diagonal (Hall) conductivity and the off-diagonal current fluctuation\cite{Takahashi}. 
Consequently, the FDT is accepted as a universal relation that holds for all components of the conductivity and the corresponding current fluctuation in classical macroscopic systems. 
This is in sharp contrast to the 
fluctuations in {\em nonequilibrium} states,
for which the FDT in the above sense does not hold in general
and the fluctuations obey other relations
\cite{
noneq_macro_1/f,
noneq_meso_quantum1,noneq_meso_quantum2,
Buttiker_FDT,Buttiker_review,Kobayashi_review,meso_exp,
noneq_meso_quantum_heat,
noneq_macro_meso_quantum,SUS1993,
noneq_macro_classical1,noneq_macro_classical2,noneq_macro_classical3,
fluctuationtheorem1_review,fluctuationtheorem2_theory,fluctuationtheorem3_exp,
glass_macro_review}.
For definiteness, we use the term `FDT' specifically for the 
above universal relation between conductivity and 
{\em equilibrium} current fluctuation.

The most important aspect of the FDT is that it connects different and independent experiments. 
The conductivity is obtained via the measurement of electric current in a nonequilibrium state, while the current fluctuation via that of the time correlation of electric currents in an equilibrium state. 
The FDT enables us to estimate the magnitude of current noises (fluctuations) from the observed conductivity and vise versa\cite{electricalengineering1,electricalengineering2}. 
Moreover, one can also estimate temperature from the observed noise and conductivity\cite{thermometer}. 
The FDT has thus been utilized widely. 

The validity of the FDT for classical systems is well established, as mentioned above.
By contrast, 
the validity of the FDT in a quantum system is nontrivial, 
since the disturbances by measurements were not considered in its "derivation"\cite{CallenWelton,Nakano,Kubo1,Kubo2}. 
In fact, in quantum systems, disturbance should be taken into account when calculating time correlation (which is the fluctuation in the FDT), and the time correlation depends strongly on the way it is measured\cite{QMT1,QMT8,QMT2,QMT7,QMT5,QMT4}. 
For example, when the measuring apparatus is rather violent in the sense that it absorbs electromagnetic quanta of the conductor, the FDT is violated at high  frequencies, $\hbar \omega \gg k_{\rm B} T$\cite{QMT7}.
This type of violation occurs because 
such an apparatus cannot measure the zero-point fluctuation which dominates the high-frequency fluctuation\cite{QMT1,QMT8,QMT2,QMT7}. 

Then, two questions were raised by Refs.\ \onlinecite{FS2016,SF2017}.
One is whether the FDT can be violated even at low frequencies, $\hbar \omega \ll k_{\rm B} T$ (particularly at $\omega=0$).
The other is a more fundamental question: 
Does the FDT always hold when fluctuation is measured 
in a way free from avoidable disturbances, i.e., 
in an ideal way that simulates classical ideal measurements as closely as possible?
Such an ideal quantum measurement is called a 
{\em quasi-classical measurement}
(see Refs.\ \onlinecite{FS2016,SF2017} 
for its rigorous definition). 

We henceforth focus on the fluctuations in the bulk region of a macroscopic conductor, not on the fluctuation in the reservoirs connected to the mesoscopic conductor \cite{Buttiker_FDT,Buttiker_review,Kobayashi_review,meso_exp}.
 (See Sec.\ \ref{sec:note} for related discussion.)
We expect a universal property,
which is independent of microscopic details,
 to appear in the thermodynamic limit. 
In fact, Refs.\ \onlinecite{FS2016,SF2017} proved the universal result that the FDT is partially violated in macroscopic systems even when fluctuation is measured by a quasi-classical measurement and even at $\omega=0$ \cite{FS2016,SF2017}. 
Care is needed for the origin of the violation.
Since the FDT here is about equilibrium fluctuations, 
this violation is never caused by a nonequilibrium effect.
It is caused by a genuine quantum effect, 
which appears even on a macroscopic scale.

Unfortunately, 
Refs.\ \onlinecite{FS2016,SF2017} did not refer to the magnitude of the violation in realistic physical systems. 
In the previous paper\cite{KAS2018}, 
we calculated the magnitude of the violation of the FDT at $\omega=0$ for the conductivities and current fluctuations in a two-dimensional electron system. 
We found that the FDT is significantly violated for the off-diagonal conductivity and 
fluctuation, while it holds for the diagonal conductivity and fluctuation.
The magnitude of the violation can even overwhelm that of the off-diagonal (Hall) conductivity itself at low temperatures and in strong magnetic fields.
Unlike the conventional belief based on the FDT, the off-diagonal fluctuation and conductivity obey entirely different physics.
Still, we need more elaborate calculations to clarify this point due to the limitation of the self-consistent Born approximation (SCBA)\cite{SCBA} we employed\cite{KAS2018} (See Sec.\ \ref{sec:result} for details). 

In this paper, we study the off-diagonal current fluctuation $S_{xy}$ in the bulk region of a two-dimensional macroscopic conductor in a strong magnetic field, assuming that the measurement is 
quasi-classical, i.e., 
it is made in an ideal way that simulates classical ideal measurements as closely as
possible\cite{FS2016, SF2017}. 
By rigorous analytic and exact numerical calculations, we clarify the physics bringing the difference between the off-diagonal current fluctuation, $S_{xy}$, and the off-diagonal (Hall) conductivity, $\sigma_{xy}$. 
We also clarify the behavior of $S_{xy}$ as a function of the filling factor in the presence of random impurity potential. 

We first show that not only extended states but also localized states contribute to $S_{xy}$, even at low temperatures. 
This result contrasts with the fact that the localized states do not affect $\sigma_{xy}$\cite{AokiAndo}, and also apparently contradicts the naive expectation from the FDT that connects the current fluctuation and conductivity directly.
We further find that $S_{xy}$ is insensitive to the impurity concentration and thus increases almost linearly as a function of the Landau level filling factor.
As an application of this property, we propose a method of estimating the electron density from the observed off-diagonal fluctuation.

This paper is organized as follows. 
In Sec.\ \ref{sec:FDTviolation}, we briefly explain the results of the FDT violation.
In Sec.\ \ref{sec:contribution}, by analytic calculations, we rigorously show that all states including localized states contribute to $S_{xy}$. 
In Sec.\ \ref{sec:pure}, we briefly explain the physics of the two-dimensional electron system in a perpendicular magnetic field, 
and rigorously calculate the off-diagonal current fluctuation of the pure system. 
In Sec.\ \ref{sec:modelmethod}, we explain a model and a method of our numerical calculation.
In Sec.\ \ref{sec:result}, we show that the off-diagonal current fluctuation is almost proportional to the filling factor, whose reason is discussed in Sec.\ \ref{sec:discussions}. 
In Sec.\ \ref{sec:application}, we propose a new method of estimating the filling factor and the electron density from the observed off-diagonal current fluctuation.  
Notes for observing our results in practical experiments are given in Sec.\ \ref{sec:note}. 
Finally, we summarize our study in Sec.\ \ref{sec:summary}.

\section{FDT Violation and properties of current fluctuation}
\label{sec:FDTviolation}

\subsection{Dependence on the way of measurement}

In general, temporal fluctuation is characterized by the spectral intensity, the Fourier transform of the temporal correlation function.
The observation of time correlation needs successive pairs of measurements,  
and 
the preceding measurement affects the outcome of the later one. 
Consequently, the observed fluctuation depends on the way of measurements
\cite{QMT1,QMT8,QMT2,QMT7,QMT4,QMT5}.

The importance of this fact was first emphasized in quantum optics. 
For example, when the photon correlation is measured by a photon-absorbing detector
the observed correlation is the normal order product of the field operators\cite{QMT1}.
As a result, the zero-point fluctuation cannot be measured 
by such a detector\cite{QMT1,QMT2,QMT7}.
On the other hand, when the quantum counter is used as a detector 
the observed correlation is the anti-normal order product\cite{QMT8}, 
and the zero-point fluctuation {\em can} be measured\cite{QMT2,QMT7}.
Another operator is obtained when another type of detector, 
such as quantum non-demolition detectors\cite{QNDpra,QND1}, is used.

The same can be said for the current fluctuation in electrical conductors because 
the current is a bosonic excitation like the photon field.
For instance, many works on mesoscopic conductors showed that 
the zero-point fluctuation cannot be measured by detectors which 
absorb quanta in the circuit\cite{VM1,VM2,VM3,VM4,VM5,VM6,VM7}.
This corresponds to the famous result by Glauber for the photon field \cite{QMT1}.
Another result is obtained when another type of detector is used\cite{Koch1982,QMT7}.

\subsection{Universal results for equilibrium fluctuation in macroscopic systems}

From the above-mentioned results of the previous works,
it seemed believed  
that no universal results would exist for what operator is measured when 
fluctuation is measured.
For {\em equilibrium} fluctuations in {\em macroscopic} systems, however, 
Ref.~\onlinecite{FS2016} proved rigorously that a universal result does exist:
As long as the measurement is quasi-classical, the observed time correlation of 
a macroscopic system 
is {\em always} a symmetrized one.

Remarkably, this result holds for 
{\em all} macroscopic quantum systems and 
for {\em all} quasi-classical measurements, 
i.e., for all 
measurements that simulate the ideal classical measurement as closely as possible\cite{FS2016,SF2017}.

We are interested in such an ideal case 
where the 
equilibrium fluctuation in a macroscopic system is measured quasi-classically.
Then, as a consequence of the above universal result \cite{FS2016,SF2017,KAS2018}, 
the FDT is violated, as will be summarized below.

%

\subsection{Two-dimensional macroscopic system}

We apply the above result of Refs.~\onlinecite{FS2016,SF2017} to 
a two-dimensional macroscopic system.

The spectral intensity of equilibrium current fluctuation 
observed through any quasi-classical measurement is given by 
\begin{align}
	S_{\mu\nu}(\omega):=
	\frac{1}{\Omega}
	 \int_{0}^{\infty}
	\langle\frac{1}{2}\{\hat{J}_{\nu}(0), \hat{J}_{\mu}(t)\}\rangle_{\rm{eq}}
	e^{i \omega t}dt, 
	\label{eq:definitionofS}
\end{align}
where $\{ \hat{A}, \hat{B} \} := \hat{A}\hat{B}+\hat{B}\hat{A}$, $\langle \bullet \rangle_{\rm eq}$ denotes the equilibrium expectation value, 
$\hat{J}_{\mu}=-e\sum_{i}^{N_e}\dot{\hat{r}}_{\mu}\ (\mu=x,y)$ is the total current with electron charge $-e$ and the number of electrons $N_e$, 
and $\Omega=L^2$ is the area of the system with size length $L$. 

On the other hand, the electric conductivity is given by the Kubo formula\cite{Kubo1, Kubo2} as
\begin{align}
	\sigma_{\mu\nu}(\omega)
	:=
	\frac{1}{\Omega}
	\int_{0}^{\infty}
  	\beta
	\langle
		\hat{J}_\nu(0);\hat{J}_\mu(t)
	\rangle
  	e^{i\omega t}dt,
	\label{eq:definitionofsigma}
\end{align}
in terms of the inverse temperature, $\beta$, and the canonical time correlation,
\begin{align}
	\langle
		\hat{A};\hat{B}
	\rangle
	:=
	\frac{1}{\beta}
	\int_{0}^{\beta}
	\left\langle
	e^{\lambda\hat{H}_{\rm eq}}
	\hat{A}
	e^{-\lambda\hat{H}_{\rm eq}}	
	\hat{B}
	\right\rangle_{\rm eq}
	d\lambda,
\end{align}
where $\hat{H}_{\rm eq}$ is the equilibrium Hamiltonian. 

In the classical limit $\hbar\rightarrow 0$, 
the symmetrized time correlation in Eq.~(\ref{eq:definitionofS}) 
and the canonical time correlation in Eq.\ (\ref{eq:definitionofsigma}) 
reduce to the same classical time correlation, leading to 
\begin{align}
\mbox{classical case: }
\beta S_{\mu\nu}(\omega) = \sigma_{\mu\nu}(\omega)
\quad \mbox{for all $\mu, \nu, \omega$},
\label{eq:FDTclassical}
\end{align}
which is called the fluctuation-dissipation theorem (FDT)\cite{Takahashi}.
From this relation, one often naively expect that $\beta S_{\mu\nu}(\omega)$ would have the same properties as $\sigma_{\mu\nu}(\omega)$. 

In quantum systems, however, the symmetrized time correlation deviates from the canonical one, and thus the FDT is violated in part, as shown in  Refs.~\onlinecite{FS2016,SF2017, KAS2018}.
To be specific, let us focus on the DC limit $\omega\rightarrow 0$, where we can capture the physics most clearly, introducing two static limit quantities, 
\begin{align}
	\beta S_{\mu\nu}:=\beta S_{\mu\nu}(0), 
	\
	\sigma_{\mu\nu}:=\sigma_{\mu\nu}(0). 
\end{align}
For the diagonal component the FDT holds,
\begin{align}
	\beta S_{xx}
	&=\sigma_{xx},
\end{align}
even in quantum systems.
{\it That is, $\beta S_{xx}$ has the same properties and the equivalent information as $\sigma_{xx}$,
as long as the measurement is quasi-classical. }
For example, only the extended states lying near the chemical potential contribute to $\sigma_{xx}$ and $\beta S_{xx}$\cite{SCBA, expofsigmaxx}. 

For the off-diagonal component, in contrast, the FDT is violated\cite{FS2016,SF2017,KAS2018},
\begin{align}
	\beta S_{xy}
	&\neq\sigma_{xy},
\end{align}
because of strong quantum effects.
The magnitude of the violation can be macroscopically large\cite{KAS2018}. 
For example, in a two-dimensional macroscopic system with a strong magnetic field at low temperature the violation approaches \cite{KAS2018}
\begin{align}
	\frac{|\sigma_{xy}-\beta S_{xy}|}{|\sigma_{xy}|}\sim \frac{\beta\hbar\omega_{\rm c}}{2}
\end{align}
with increasing $\hbar\omega_{\rm c}/\Gamma$ and $\beta\hbar\omega_{\rm c}$, where $\omega_{\rm c}$ is the cyclotron frequency and $\Gamma$ is the width of Landau levels.

Such a significant violation shows that $\beta S_{xy}$ and $\sigma_{xy}$ obey different physics in quantum systems. 
In other words, they must have different properties, in contrast to the naive expectation from the classical FDT, Eq.~(\ref{eq:FDTclassical}). 
While the properties of $\sigma_{xy}$ were extensively studied\cite{vonKlitzing,wakabayashikawaji1,wakabayashikawaji2,expofsigmaxx,SCBA,Andoxy,TKNN,Kohmoto,NiuThouless,Halperinedge,Hadju2,OnoKramer,OnoOhtsuki,OhtsukiOno,laughlinIQHE,AokiAndo}, little is known about $\beta S_{xy}$.
We will explore $\beta S_{xy}$ and prove that it has a property significantly different 
from that of $\sigma_{xy}$.

\section{contribution of individual state to $\beta S_{xy}$ and $\sigma_{xy}$}
\label{sec:contribution}
To study the properties of $\beta S_{xy}$, we consider the non-interacting two-dimensional electron system
in the $x$-$y$ plane in a uniform magnetic field $B$ along the $z$-axis. 
The Hamiltonian for a single electron reads 
\begin{align}
\hat{H}
=
\frac{\hat{\bm{\pi}}^2}{2m}
+ V(\hat{\bm{r}}),
 \label{eq:Hamiltonian}
\end{align}
where $m$ is the electron mass, $\hat{\bm{\pi}}=\dot{\hat{\bm r}}=\hat{\bm{p}}+e\bm{A}(\hat{\bm{r}})$ is a dynamical momentum with electron charge $-e$ and vector potential $\bm{A}$, and $V(\hat{\bm{r}})$ denotes the impurity potential.

For simplicity, we assume that the system is invariant under rotation by $\pi/2$ about the $z$-axis. 
Then we can write 
\begin{align}
	\beta S_{xy}&=\frac{\beta S_{xy}-\beta S_{yx}}{2}, 
	\label{eq:Sxy}
	\\
	\sigma_{xy}&=\frac{\sigma_{xy}-\sigma_{yx}}{2}. 
	\label{eq:sigmaxy}
\end{align}
After a straightforward calculation, 
we obtain
\begin{align}
	\beta S_{xy}
	&=
	\sum_{\alpha}
	f_{\beta, \mu}(E_{\alpha})
	D_{\alpha}, 
	\label{eq:betaSxyleaman}
	\\
	D_{\alpha}
	&=\frac{\hbar e^2}{i\Omega m^2}\sum_{\alpha'(\neq \alpha)}M_{\alpha\alpha'}I_\beta\left(\frac{E_{\alpha'}-E_{\alpha}}{\hbar}\right), 
	\label{eq:Dalpha}	
	\\
	M_{\alpha\alpha'}&=\frac{
		\bra{\alpha}\hat{\pi}_x\ket{\alpha'}\bra{\alpha'}\hat{\pi}_y\ket{\alpha}
		-\bra{\alpha}\hat{\pi}_y\ket{\alpha'}\bra{\alpha'}\hat{\pi}_x\ket{\alpha}
		}
		{
			(E_{\alpha}-E_{\alpha'})^2
		}
	\\
	I_{\beta}(\omega)
	&=
	\frac{\beta\hbar\omega}{2}\coth \frac{\beta\hbar\omega}{2}\sim
	\begin{cases}
		1 & (\beta \hbar |\omega| \ll 1),\\
		\beta \hbar |\omega| /2 & (\beta \hbar |\omega| \gg 1),
	\end{cases}
	\label{eq:I_beta}
\end{align}
where $\ket{\alpha}$ is an energy eigenstate with eigenenergy $E_{\alpha}$, and $f_{\beta, \mu}(x)$ is the fermi-distribution function with inverse temperature $\beta$ and chemical potential $\mu$.
This expression resembles the well-known form of $\sigma_{xy}$\cite{AokiAndo,TKNN}:
\begin{align}
	\sigma_{xy}
	&=
	\sum_{\alpha}
	f_{\beta, \mu}(E_{\alpha})
	C_{\alpha}, 
	\label{eq:Aokiando}\\
	C_{\alpha}
	&=
	\frac{\hbar e^2}{i\Omega m^2}
	\sum_{\alpha'(\neq \alpha)}M_{\alpha\alpha'}
	\label{eq:Calpha}, 
\end{align}
except for the factor $I_{\beta}$ in Eq.~(\ref{eq:Dalpha}).
We can confirm that the FDT holds in the classical limit $\hbar\rightarrow 0$, in which $I_{\beta}\rightarrow 1$. 

For a localized state, $\ket{\alpha}$, we have 
\begin{align}
	\bra{\alpha}\hat{\pi}_{\mu}\ket{\alpha'}=\frac{m}{i\hbar}\bra{\alpha}\hat{r}_{\mu}\ket{\alpha'}{(E_{\alpha'}-E_{\alpha})}.
	\label{eq:Heisenberg}
\end{align}
This equality and $\sum_{\alpha'}\ket{\alpha'}\bra{\alpha'}=1$ leads to $C_{\alpha}=0$\cite{AokiAndo}, implying that the localized state cannot contribute to $\sigma_{xy}$. 
A similar argument also shows that the localized states affect neither to the diagonal conductivity $\sigma_{xx}$ nor to the diagonal current fluctuation $\beta S_{xx}$. 

In contrast, this argument is not applicable to $D_{\alpha}$ owing to the factor, $I_{\beta}$, which describes a genuine quantum effect \cite{FS2016,SF2017,KAS2018}. 
Namely, all states {\em including localized states} contribute to $\beta S_{xy}$, unlike the case of 
$\sigma_{xy}$. 
In the following sections, we will quantitatively demonstrate this property for disordered systems where almost all states are localized\cite{Anderson,scalingtheory,Andothouless} by showing that $D_\alpha$ of each localized state is of the same order as that of each extended state.

\section{pure system}
\label{sec:pure}

In the following sections, we take the Landau gauge $A(\bm{r})=(0,Bx)$, and consider a finite size $(L\times L)$ system with periodic boundary conditions in both directions. 

As a reference system, we first consider a pure system ($V(\bm{r})=0$), where all states are extended. 
We take the simultaneous eigenstates $\ket{NX}$ of the Hamiltonian and the guiding center coordinate as $\ket{\alpha}$\cite{Andothouless}. 
Their eigenvalues are given as
\begin{align}
	E_{NX}&=\hbar\omega_{\rm c}\left(N+\frac{1}{2}\right), 
	\label{eq:ENX}
\\
	X&=j\Delta X\ (j=1,2,\cdots,N_{\phi}),
\end{align}
respectively, where $\omega_{\rm c}=eB/m$ is the cyclotron frequency, $N=0,1,2,\cdots$ is the index of Landau levels, $N_{\phi}=\Omega/2\pi l^2$ is the number of states in each Landau level, $l=\sqrt{\hbar/eB}$ is the magnetic length, and $\Delta X=2\pi l^2/L$ is the pitch width of the discretized guiding center coordinate. 
The eigenfunction is also given as 
\begin{align}
	\braket{\bm{r}|N, X}=\sum_{M\in \mathbb{Z}}&\frac{1}{\sqrt{L}}\chi_{N}(x-X-ML)\notag\\
	&\times\exp\left(-i\frac{(X+ML)y}{l^2}\right), 
	\label{eq:basis}
\end{align}
where we introduce
\begin{align}
	\chi_{N}(x)&=\frac{1}{\sqrt{2^NN!\sqrt{\pi}l}}H_N\left(\frac{x}{l}\right)\exp\left(-\frac{1}{2}\left(\frac{x}{l}\right)^2\right),
\end{align}
using the Hermite polynomial, $H_N(x)$. 

Using Eqs. (\ref{eq:betaSxyleaman})-(\ref{eq:I_beta}), (\ref{eq:ENX}) and 
	\begin{align}
		&\bra{NX}\hat{\pi}_{x}\ket{N'X'}
		\nonumber\\
		&=\frac{lm\omega_{\rm c}}{\sqrt{2}}(\sqrt{N+1}\delta_{N+1,N'}+\sqrt{N}\delta_{N-1,N'})\delta_{X,X'}, 
		\label{eq:pix}
		\\
		&\bra{NX}\hat{\pi}_{y}\ket{N'X'}
		\nonumber\\
		&=\frac{lm\omega_{\rm c}}{\sqrt{2}}i(\sqrt{N+1}\delta_{N+1,N'}-\sqrt{N}\delta_{N-1,N'})\delta_{X,X'}, 
		\label{eq:piy}
	\end{align}
we can express $\beta S_{xy}$ of the pure system as
\begin{align}
	\beta S^{(\rm pure)}_{xy}
	=-\frac{e^2}{h}
	I_{\beta}(\omega_{\rm c}) \nu,
	\label{eq:Spure}
\end{align}
in terms of the the Landau level filling factor, $\nu=N_{e}/N_\phi=2\pi l^2n$, and the electron density, $n=N_e/\Omega$. 
Although Eq. (\ref{eq:Spure}) resembles the well-known form of $\sigma_{xy}$ of the pure system, 
\begin{align}
	\sigma^{(\rm pure)}_{xy}
	=-\frac{e^2}{h} \nu, 
\end{align} 
$\beta S^{(\rm pure)}_{xy}$ is $I_{\beta}(\omega_{\rm c})$ times larger than $\sigma^{(\rm pure)}_{xy}$. 
Therefore, the FDT is violated. 

In pure systems, both $S_{xy}$ and $\sigma_{xy}$ are proportional to $\nu$ since each state is extended and equally contributes to both of them.
However, in disordered systems, the majority of states get localized.
We should thus pay attention to the difference between the extended and localized states and how they affect these two quantities. 

\section{model of disordered system and method of numerical calculation} 
\label{sec:modelmethod}

Now, let us consider the disordered systems. 
For simplicity, we employ short-range impurity potential, 
\begin{align}
	V(\hat{\bm{r}})=\sum_{i}u_i\delta(\hat{\bm{r}}-\bm{R}_i),
\end{align}
where $\bm{R}_{i}=(x_i, y_i)$ is the position of the $i$th scatterer and $u_{i}$ is its strength. 
They are distributed uniformly in the regions 
$x_{i},y_{i} \in [0,L]$ and $u_i \in [-u/2,u/2]$, respectively. 
Impurity scatterings broaden the Landau levels, and all states get localized except for few extended ones lying in the vicinity of the center of the Landau levels\cite{Andothouless}. 

We study the filling factor dependence of $\beta S_{xy}$ numerically. 
To satisfy the periodic boundary conditions, we employ $\ket{N,X}$ defined by Eq. (\ref{eq:basis}) as basis kets.
It is straightforward to derive the analytical expression of the Hamiltonian matrix elements, $\braket{N,X|\hat H|N',X'}$. 
Then, we can obtain all eigenpairs, $E_\alpha$ and $|\alpha\rangle$, by the numerical diagonalization, and can also evaluate $\beta S_{xy}$ and $\sigma_{xy}$ using Eq,~(\ref{eq:betaSxyleaman}) and (\ref{eq:Aokiando}).

The values of the parameters are taken as follows. 
We are interested in the low-temperature and strong-magnetic-field regime such that 
\begin{align}
	\beta\hbar\omega_{\rm c} \gg 1, 
	\label{eq:condition1}\\
	\epsilon:=\Gamma/\hbar\omega_{\rm c}<1,  
	\label{eq:condition2}
\end{align}
in which the FDT is macroscopically violated\cite{KAS2018}.  
Here, we introduced the energy scale: 
\begin{align}
	\Gamma
	=\sqrt{4\frac{n_{i}u^2}{2\pi l^2}}
	=\sqrt{\frac{2}{\pi}\frac{\hbar}{\tau}\hbar\omega_{\rm c}}. 
\end{align}
This is the asymptotic form of the Landau level width at $\omega_{\rm c}\tau\rightarrow \infty$ \cite{SCBA}, where $\tau$ is the scattering time at $B=0$ and $n_{i}$ is the impurity concentration. 
To satisfy Eqs.~(\ref{eq:condition1}), (\ref{eq:condition2}), we take $\beta\hbar\omega_{\rm c}=250$, $\epsilon=0.1, 0.2, 0.3, 0.4$. 
To avoid peculiarity of $\delta$-potential scatters\cite{Andothouless,Andoimpuconcentration}, we have to take $n_{\rm i}$ high enough such that $2\pi l^2 n_{i}>1$.
We thus take $2\pi l^2 n_{\rm i}=7$.
For a reasonable size of computation, we choose $N_{\phi}=10000$ and restrict the range of $N$ as  $N\in[0,2]$. 

\section{numerical results}
\label{sec:result}
\begin{figure}[t]
\includegraphics[width=8cm,pagebox=cropbox,clip]{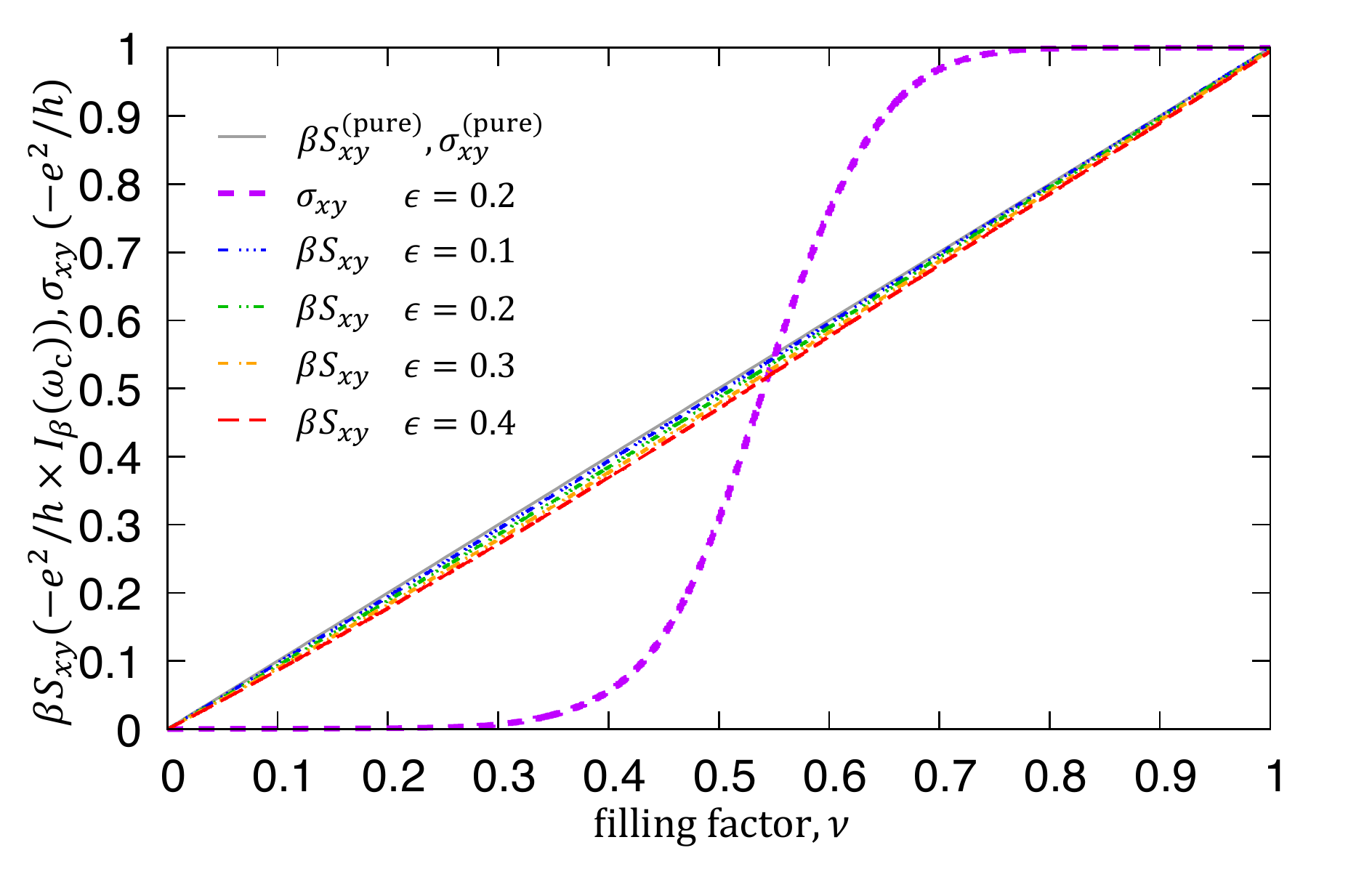}
\caption{
	(color online). 
	$\beta S_{xy}$ as a function of the filling factor $\nu$ for 
$\epsilon=0.1$-$0.4$ at $k_{\rm B}T=\hbar\omega_{\rm c}/250$.  
	As a reference, 
	$\sigma_{xy}$ for $\epsilon=0.2$, 
	$\beta S_{xy}^{({\rm pure})}$, and 
	$\sigma_{xy}^{({\rm pure})}$ at the same temperature 
	are also plotted. 
	}
\label{fig:vsfilling}
\end{figure}
\begin{table}[t]
  \centering
  \begin{tabular}{|c|c|c|c|} \hline
    $\epsilon$ & $R^2$ & $a(-e^2/h\times I_{\beta}(\omega_{\rm c}))$ & $\delta a$  \\ \hline
    0.1 & 0.9997 & $0.9945\pm0.0001$ & $0.006$  \\ \hline
    0.2 & 0.9987 & $0.9877\pm0.0001$ & $0.013$  \\ \hline
    0.3 & 0.9969 & $0.9797\pm0.0002$ & $0.020$  \\ \hline
    0.4 & 0.9945 & $0.9720\pm0.0002$ & $0.028$  \\ \hline
  \end{tabular}
  \caption{
  $\epsilon=\Gamma/\hbar\omega_{\rm c}$ is system parameter
  and $R^2$ is the coefficient of the determination, 
  which is the index of the accuracy of the linear fitting. 
  $\delta a$ is the relative error between 
the slope $a$ of fitting function and that of $\beta S_{xy}^{({\rm pure})}$.} 	
\label{fig:table}
\end{table}
Figure \ref{fig:vsfilling} shows the 
numerical results of $\beta S_{xy}$ and $\sigma_{xy}$ as a function of
the filling factor $\nu$. 
The units of the vertical axis are $-e^2/h\times I_{\beta}(\omega_{\rm c})$ and $-e^2/h$, respectively, 
so that they can be plotted
on the same scale. 

The $\nu$ dependence of $\sigma_{xy}$ is far from the linear one of $\sigma_{xy}^{(\rm pure)}$, showing quantized plateaus of $\sigma_{xy}=0$ and $-e^2/\hbar$ at around $\nu=0$ and $\nu=1$, respectively, which is nothing but the integer quantum Hall effect\cite{vonKlitzing,wakabayashikawaji1,wakabayashikawaji2,expofsigmaxx,SCBA,Andoxy,TKNN,Kohmoto,NiuThouless,Halperinedge,Hadju2,OnoKramer,OnoOhtsuki,OhtsukiOno,laughlinIQHE,AokiAndo}.
These plateaus are attributable to the fact that the localized state does not contribute to $\sigma_{xy}$ and that all states are localized except for few ones near the Landau level center, as mentioned in Sec.\ref{sec:contribution}.
In our calculation, the finite temperature and finite-size effects slightly smear the plateau structures.

In contrast, $\beta S_{xy}$ grows almost linearly with increasing $\nu$.
This $\nu$ dependence is qualitatively different from that of Ref.\ \onlinecite{KAS2018}, 
while their orders of magnitude are the same (in particular, their values are precisely equal at $\nu=1$). 
This discrepancy is caused by SCBA \cite{SCBA} employed there, which can not describe the Anderson localization. 

We can fit the $\nu$ dependence of $\beta S_{xy}$ with a regression line with high accuracy, as shown in Tab.~\ref{fig:table}.
For all values of $\epsilon$ listed in the table, the proportionality coefficient, $a$, is almost equal to $-e^2/h\times I_\beta(\omega_{\rm c})$.
The coefficient of determination $R^2$ is almost equal to $1$. 
The relative error, $\delta a$, of the slope of $S_{xy}$ from $a$ is within the order of $10^{-3}$ to $10^{-2}$. 
From these results, we conclude that the behavior of $\beta S_{xy}$ is almost unchanged by impurity scatterings, in contrast to $\sigma_{xy}$.
Localized states and extended states equally contribute to $\beta S_{xy}$: $D_{\alpha}$ of each state is of the same order of magnitude.

\section{Physical origin}
\label{sec:discussions}

In this section, we discuss why $\beta S_{xy}$ is much more insensitive to the impurities than $\sigma_{xy}$.
As in the previous sections, we consider the low-temperature and strong-magnetic-field regime where Eq.\ (\ref{eq:condition1}) and (\ref{eq:condition2}) are satisfied.
We assume the chemical potential lies in the $N$-th Landau level.

\subsection{$\sigma_{xy}$ in terms of $\Im\sigma_{xy}(\omega)$}
To discuss $\beta S_{xy}$, we first express $\sigma_{xy}$ ($=\Re \sigma_{xy}(0)$) in terms of $\Im\sigma_{xy}(\omega)$ as
\begin{align}
	\sigma_{xy}=\mathcal{KK}[\Im\sigma_{xy}(\omega)], 
	\label{eq:KKsigma}
\end{align}
where
\begin{align}
	\mathcal{KK}[A (\omega)]=\int_{-\infty}^{\infty}\frac{d\omega}{\pi}\frac{\mathcal{P}}{\omega}A (\omega)
\end{align}
stands for the Kramers-Kroning transformation of $A(\omega)$\cite{KTH}. 

In the pure system, the integrand $\Im\sigma_{xy}(\omega)$ has sharp peaks at $\omega=\pm\omega_{\rm c}$ as
\begin{align}
	 \Im\sigma_{xy}^{(\rm pure)}(\omega)=-\frac{e^2\omega_{\rm c}^2}{4\omega}\nu(\delta(\hbar\omega+\hbar\omega_{\rm c})+\delta(\hbar\omega-\hbar\omega_{\rm c})). 
\end{align}
In a disordered system, they are broadened and split into many structures centered at around $\omega = 0, \pm \omega_{\rm c}, \pm 2 \omega_{\rm c}, \ldots$, with width of $\mathcal{O}(\Gamma/\hbar)$.
We decompose the spectra, $\Im\sigma_{xy}(\omega)$, into the intra- and the inter-Landau level contributions, 
\begin{align}
\Im\sigma_{xy}^{\rm intra}(\omega)&=\Im\sigma_{xy}(\omega)\theta(\omega_{\rm c}/2-|\omega|)
\label{eq:sigma_decmposed1}\\
\Im\sigma_{xy}^{\rm inter}(\omega)&=\Im\sigma_{xy}(\omega)\theta(|\omega|-\omega_{\rm c}/2))
\label{eq:sigma_decmposed2}
\end{align}
using the Heaviside step function, $\theta(x)$.
Similarly, we decompose the deviation of $\sigma_{xy}$ from the pure case, 
\begin{align}
\Delta\sigma_{xy}:=\sigma_{xy}-\sigma_{xy}^{({\rm pure})},
\end{align}
into the two corresponding contributions,
\begin{align}
	\Delta\sigma_{xy}^{\rm intra}&:=\mathcal{KK}[\Im\sigma_{xy}^{\rm intra}(\omega)],
	\label{eq:deltasigmaintra}
\\
	\Delta\sigma_{xy}^{\rm inter}&:=\mathcal{KK}[\Im\sigma_{xy}^{\rm inter}(\omega)-\Im\sigma_{xy}^{(\rm pure)}(\omega)],
	\label{eq:deltasigmainter}
\end{align}
using Eq.~(\ref{eq:KKsigma}) and the fact that $\Im\sigma_{xy}^{\rm intra}(\omega)$ vanishes in the pure case.

\begin{figure}[t]
	\includegraphics[width=8cm,pagebox=cropbox,clip]{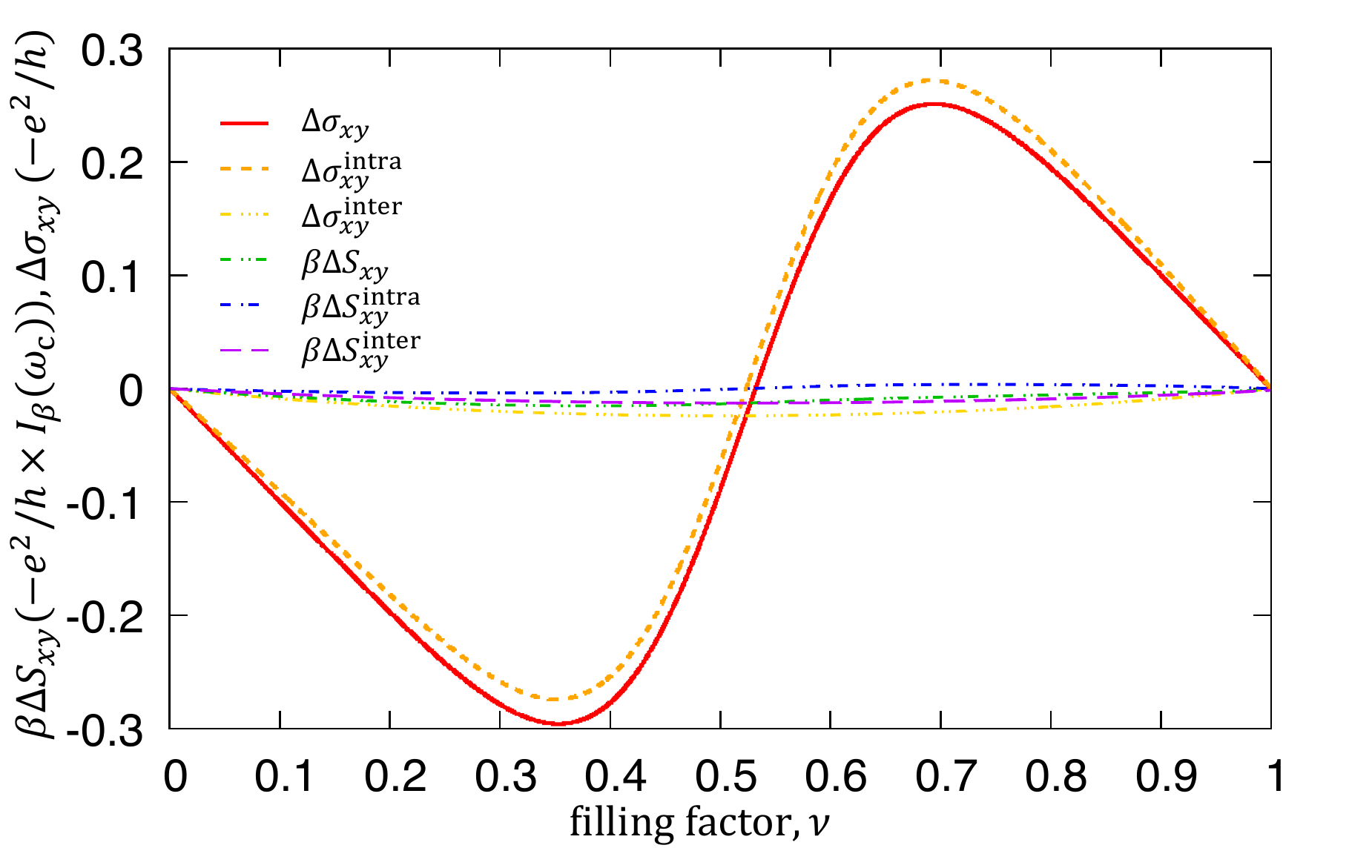}
	\caption{
	(color online).  
	$\Delta\sigma_{xy}$ and $\beta \Delta S_{xy}$ as a function of $\nu$ for $\epsilon=0.2$ at $k_{\rm B}T=\hbar\omega_{\rm c}$/250.
$\Delta\sigma_{xy}^{\rm inter}$, $\Delta\sigma_{xy}^{\rm intra}$, 
$\beta \Delta S_{xy}^{\rm inter}$, and $\beta \Delta S_{xy}^{\rm intra}$  
are also plotted. 
	}
	\label{fig:deltavsfilling}
\end{figure}

Figure \ref{fig:deltavsfilling} shows the numerical results of $\Delta\sigma_{xy}, \Delta\sigma_{xy}^{\rm intra}$ and $\Delta\sigma_{xy}^{\rm inter}$ as functions of $\nu$. 
It is seen that $|\Delta\sigma_{xy}^{\rm inter}|$ is much smaller than $|\Delta\sigma_{xy}^{\rm intra}|$.
This can be understood as follows.

\begin{figure*}[t]
\centering
	\centering
	\includegraphics[width=\textwidth]{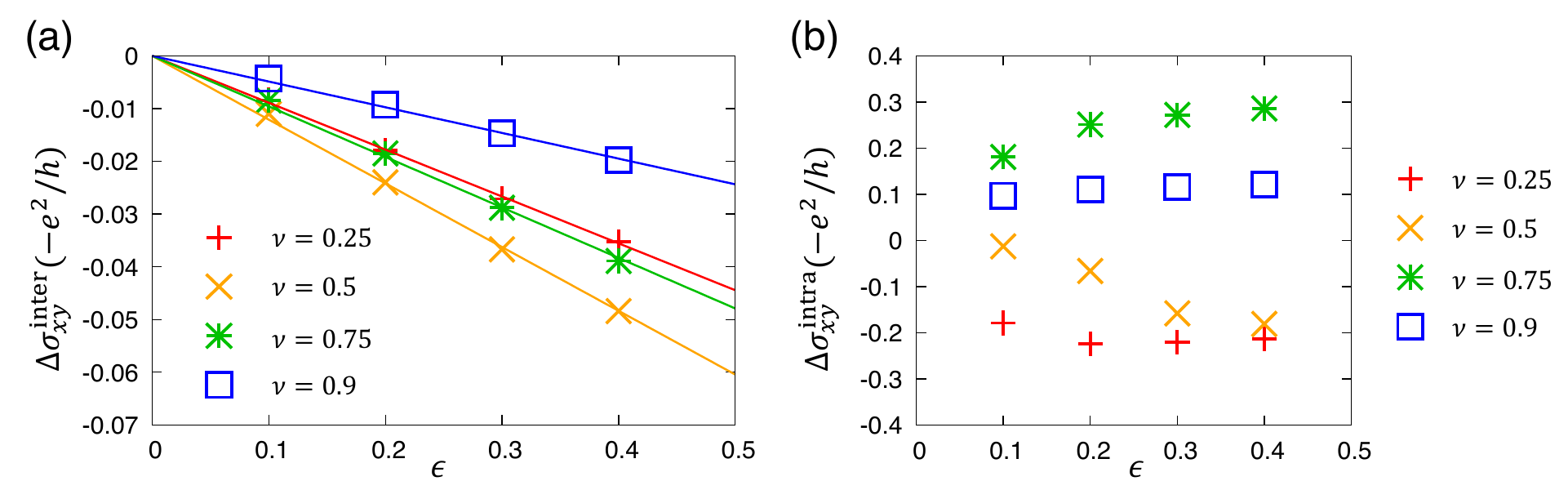}
	\caption{
	(color online).  
	(a) $\Delta\sigma_{xy}^{\rm inter} $ and (b) $\Delta\sigma_{xy}^{\rm intra}$ as a function of $\epsilon$ for 
$\nu=0.25$-$0.9$ 
at $k_{\rm B}T=\hbar\omega_{\rm c}$/250.  
The solid lines in (a) are linear fits to the data. 
Although $\Delta\sigma_{xy}^{\rm inter}$ is almost proportional to $\epsilon$, $\Delta\sigma_{xy}^{\rm intra}$ has 
contribution of order $\epsilon^0$. }
\label{fig:deltasigma}
\end{figure*}

As shown in appendix \ref{ap:inter}, by considering the broadening and loss of the spectra, we can evaluate $\Delta\sigma_{xy}^{\rm inter}$ as
\begin{align}
	\Delta\sigma_{xy}^{\rm inter}=\mathcal{O}\left(\epsilon\times \frac{e^2}{h}\right),
	\label{eq:deltasigmainter_order}
\end{align}
which is consistent with the numerical results shown in Fig.~\ref{fig:deltasigma}(a). 

In contrast, $\Delta\sigma_{xy}^{\rm intra}$ does not depend on $\epsilon$ in its lowest order, 
\begin{align}
	\Delta\sigma_{xy}^{\rm intra}=\mathcal{O}\left(\epsilon^0\times \frac{e^2}{h}\right)
	\label{eq:deltasigmaintra_order}
\end{align}
because the integer quantum Hall effect occurs (see Appendix \ref{ap:intra}).
Again, this evaluation is consistent with the numerical results shown in Fig.~\ref{fig:deltasigma}(b). 
Therefore, for small $\epsilon$, $|\Delta\sigma_{xy}^{\rm inter}| \ll |\Delta\sigma_{xy}^{\rm intra}|$ and hence $\Delta\sigma_{xy} \simeq \Delta\sigma_{xy}^{\rm intra}$.

\subsection{$\beta S_{xy}$ in terms of $\Im\sigma_{xy}(\omega)$}

\begin{figure*}[t]
\centering
	\centering
	\includegraphics[width=\textwidth]{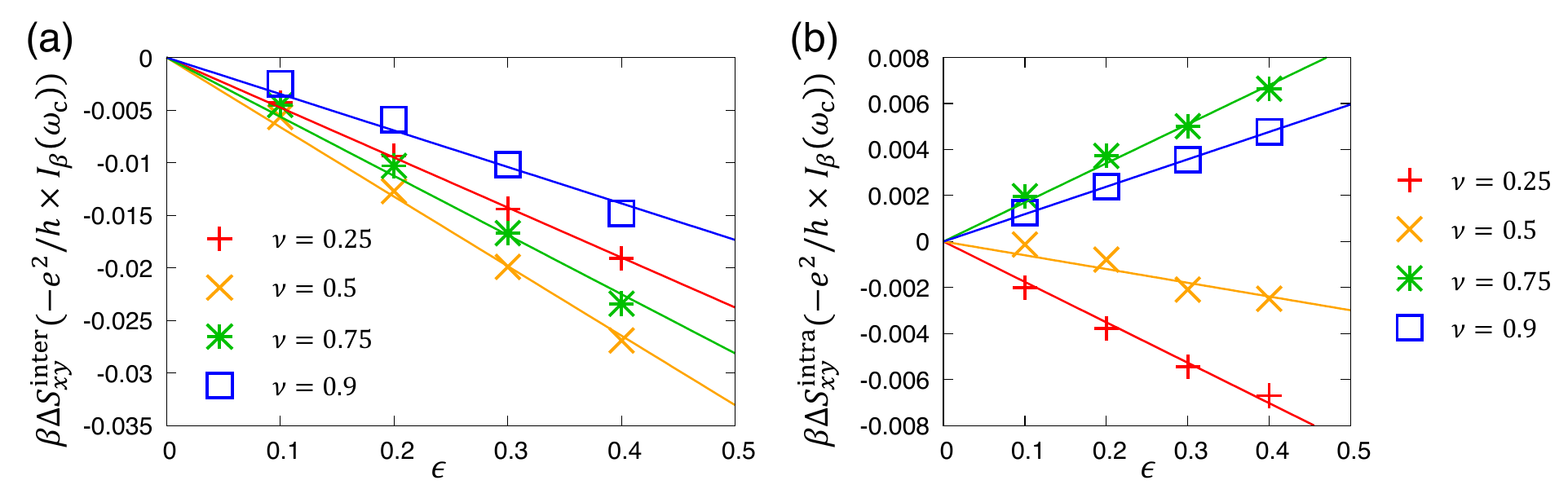}
	\caption{
	(color online).  
	(a) $\beta\Delta S_{xy}^{\rm inter} $ and (b) $\beta\Delta S_{xy}^{\rm intra}$ as a function of $\epsilon$ for 
$\nu=0.25$-$0.9$
at $k_{\rm B}T=\hbar\omega_{\rm c}$/250.  
	They are almost proportional to 
$\epsilon$, as seen from the linear fits (solid lines). 
		}
\label{fig:deltaS}
\end{figure*}

Next, we consider the off-diagonal current fluctuation. 
According to Ref.\ \onlinecite{KAS2018}, $\beta S_{xy}$ can be written as
\begin{align}
	\beta S_{xy}=\mathcal{KK}[I_{\beta}(\omega)\Im\sigma_{xy}(\omega)]. 
\end{align}
Similarly to Eqs.~(\ref{eq:deltasigmaintra}) and (\ref{eq:deltasigmainter}), 
we decompose the deviation of $\beta S_{xy}$ from the pure case,
\begin{align}
\beta\Delta S_{xy} 
&:=
\beta S_{xy}-\beta S_{xy}^{({\rm pure})}
\\
&=\mathcal{KK}[I_{\beta}(\omega)\{\Im\sigma_{xy}(\omega)-\Im\sigma_{xy}^{({\rm pure})}(\omega)\}], 
\end{align}
into the intra- and inter-Landau level contributions, 
\begin{align}
\beta\Delta S_{xy}^{\rm intra}&:=\mathcal{KK}[I_{\beta}(\omega)\Im\sigma_{xy}^{\rm intra}(\omega)],
\label{eq:deltasigmaB}
\\
\beta\Delta S_{xy}^{\rm inter}
&:=
\mathcal{KK}[I_{\beta}(\omega)\{\Im\sigma_{xy}^{\rm inter}(\omega)-\Im\sigma_{xy}^{(\rm pure)}(\omega)\}
]. 
\label{eq:deltasigmaA}
\end{align}
Since $|\Im \sigma_{xy}^{\rm inter}|$ has a dominant peak at $\omega \simeq \pm \omega_{\rm c}$, we can evaluate $\beta\Delta S_{xy}^{\rm inter}$ as
\begin{align}
\beta\Delta S_{xy}^{\rm inter}
&\simeq 
I_{\beta}(\omega_{\rm c}) \mathcal{KK}
[\Im\sigma_{xy}^{\rm inter}(\omega)
-\Im\sigma_{xy}^{(\rm pure)}(\omega)]
\nonumber\\
& = I_{\beta}(\omega_{\rm c})\Delta\sigma_{xy}^{\rm inter}
=\mathcal{O}\left(\epsilon\times \frac{e^2}{\hbar}I_{\beta}(\omega_{\rm c})\right),
	\label{eq:deltaSinter1}
\end{align}
where we have used Eq.~(\ref{eq:deltasigmainter_order}). 
This result is consistent with the numerical results in Fig.~\ref{fig:deltaS}(a). 
Since $|\Im\sigma_{xy}^{\rm intra}|$ has the highest peak at $\omega=\mathcal{O}(\Gamma/\hbar)$, we can evaluate $\beta\Delta S_{xy}^{\rm intra}$ as 
\begin{align}
\beta\Delta S_{xy}^{\rm intra}
&\simeq I_{\beta}(\mathcal{O}(\Gamma/\hbar))\Delta\sigma_{xy}^{\rm intra}
\nonumber\\
&\simeq 
\mathcal{O}(\beta \Gamma) \Delta\sigma_{xy}^{\rm intra}
=\mathcal{O}\left(\epsilon\times \frac{e^2}{h}I_{\beta}(\omega_{\rm c})\right),
	\label{eq:deltaSintra2}
\end{align}
using Eqs.\ (\ref{eq:I_beta}) and (\ref{eq:deltasigmaintra_order}), which is again consistent with the numerical results in Fig.~\ref{fig:deltaS}(b). 

From Eqs.\ (\ref{eq:deltaSinter1}) and (\ref{eq:deltaSintra2}), 
we find 
\begin{align}
	\frac{ \beta\Delta S_{xy}}{\beta S_{xy}^{\rm (pure)}}=\mathcal{O}(\epsilon),
	\label{eq:orderofdeltaS}
\end{align}
which is negligibly small when $\epsilon\ll1$. 
This contrasts with 
\begin{align}
	\frac{\Delta\sigma_{xy}}{\sigma_{xy}^{\rm (pure)}}=\mathcal{O}(\epsilon^0), 
	\label{eq:orderofdeltasigma}
\end{align}
which can be obtained from Eqs.\ (\ref{eq:deltasigmainter_order}) and (\ref{eq:deltasigmaintra_order}). 
We thus understand why $\beta S_{xy}$ is affected much more weakly by impurities than $\sigma_{xy}$.

\section{Estimation of electron density from current fluctuation}
\label{sec:application}
In classical systems, the Hall conductivity is proportional to the electron density $n$. 
This property has been utilized to estimate the electron density from the Hall coefficient\cite{carrierdensity}. 
However, in quantum systems at low temperatures and in strong magnetic fields, the Hall conductivity deviates from the classical behavior, as shown in Fig.\ \ref{fig:vsfilling}. 
There, one cannot estimate $n$ correctly, since the value of $\sigma_{xy}$ is insensitive to $n$ in a plateau region, while it depends nonlinearly on $n$ in other regions. 
Therefore, an alternative method applicable to this regime is required. 

Our results in Fig.\ \ref{fig:vsfilling} and Tab.\ \ref{fig:table} reveal that $\beta S_{xy}$ in this regime is proportional to $\nu=2\pi l^2 n$ with fairly high accuracy. 
Furthermore, its slope is insensitive to the impurity concentration and remains almost unchanged from that of the pure system, $\beta S_{xy}^{(\rm pure)}$.
Therefore, one can estimate the filling factor $\nu$ by measuring $\beta S_{xy}$.
By substituting the observed equilibrium current fluctuation $\beta S_{xy}$ into Eq.\ (\ref{eq:Spure}), one can estimate $\nu$ as 
\begin{align}
	\nu\simeq\frac{\beta S_{xy}}{-e^2/h \times I_{\beta}(\omega_{\rm c})}. 
\end{align}
The error $\delta \nu$ of this estimation is given by 
\begin{align}
	\delta\nu&=
	\frac{1}{e^2/h \times I_{\beta}(\omega_{\rm c})}
	\sqrt{
		(\Delta \beta S_{xy})^2
		}. 
\end{align}
Using Eqs.\ (\ref{eq:deltaSinter1}) and (\ref{eq:deltaSintra2}), 
we find
\begin{align}
	\delta\nu&=\mathcal{O}(\epsilon). 
\end{align}
From the numerical results shown in Fig.\ \ref{fig:deltaS}, 
it is found to be smaller by an order of magnitude than $\epsilon\simeq10^{-1}$: 
\begin{align}
	\delta\nu \simeq 10^{-2}.
\end{align}
This calculation assumes a typical sample of GaAs-AlAs heterojunctions with 
$m\sim10^{-1}m_0$ ($m_0$: free electron mass) and mobility $\mu_{e}=e\tau/m\sim 10^4 {\rm cm^2/Vs}$ in a high magnetic field, $B\sim 10{\rm T}$, and at low temperature, $T\sim 1{\rm K}$. 
In this way, one can estimate $\nu$ with high accuracy by measuring the off-diagonal current fluctuation in experiments. 

Using the obtained value of $\nu$, the electron density $n$ is estimated as $n=2\pi l^2\nu$. 
This new method of measuring $n$ will be helpful in practical applications. 

\section{Notes on experiments}
\label{sec:note}
 
Let us discuss the relevance of our results  
to experiments, and relation to 
the works on mesoscopic systems
\cite{Buttiker_FDT,Buttiker_review,Kobayashi_review}. 

We studied 
the equilibrium current fluctuation in a {\em bulk region} of a {\em macroscopic} sample.
By contrast, 
the works on mesoscopic systems
studied 
the 
current fluctuation in {\em reservoirs} connected to the mesoscopic sample, treating the sample as a scatterer, by which incident electrons from a terminal are reflected or transmitted into the others\cite{Buttiker_FDT,Buttiker_review, Kobayashi_review}.
These works and ours thus assumed different physical situations and 
studied 
the fluctuations at different positions. 

Many experiments on mesoscopic systems
were reported so far\cite{Buttiker_review, Kobayashi_review, meso_exp}. 
However, to the authors' knowledge, 
no experiments have been reported on the off-diagonal fluctuation in 
macroscopic systems
(while ones about diagonal fluctuation were reported\cite{Koch1982}). 
A comparison of the experimental results done for both situations is desired. 

In order to measure the current fluctuation in the bulk region, one may use a device that can measure the current in a macroscopic sample without attaching physical contacts to it. 
Since there are various types of such devices, including the current transformer and the Rogowski coil\cite{currentmeasurementreview}, we expect it possible to measure fluctuation in the bulk region. 

There are various ways to realize quasi-classical measurements because they are general measurements that satisfy the conditions given in Refs.\ \onlinecite{FS2016,SF2017}.
For example, such measurements may be possible by using the heterodyning technique \cite{Koch1982}.

\section{Summary}
\label{sec:summary}
We have studied the properties of the 
equilibrium off-diagonal current fluctuation in a bulk region of a macroscopic system when the fluctuation is measured in an ideal way that simulates classical ideal measurements as closely as possible. 
It is rigorously shown that all states, including localized states, contribute to $S_{xy}$, in contrast to $\sigma_{xy}$. 
Moreover, by exact numerical calculations assuming a typical disordered two-dimensional electron system in a high magnetic field at low temperature, we find that $S_{xy}$ is insensitive to the impurity concentration.
Consequently, it increases almost linearly with the filling factor of the Landau level.
Using this novel property, we have proposed a new method of estimating electron density from the observed $\beta S_{xy}$. 
These results will help to understand the FDT violation and transport phenomena and will be useful for practical applications. 

\begin{acknowledgments}
We thank C. Urano and N. Kaneko for discussions.
This work was supported by The Japan Society for
the Promotion of Science, KAKENHI No. 19J21816,
17K05497, 21K03440, 
and 19H01810.
\end{acknowledgments} 

\appendix
\section{Order of magnitude of $\Delta\sigma_{xy}^{\rm inter}$}
\label{ap:inter}
To discuss the order of magnitude of $\Delta\sigma_{xy}^{\rm inter}$, we consider $\Im\sigma_{xy}^{\rm inter}(\omega)$ with reference to $\Im\sigma_{xy}^{\rm (pure)}(\omega)$. 
There are the following two types of deviation. 

\subsection{Broadening of the spectrum}
One is the broadening of the spectrum. 
Firstly, 
we consider the case where $\Im\sigma_{xy}^{\rm inter}(\omega)$ is broadened symmetrically as
\begin{align}
	&\Im\sigma_{xy}^{\rm inter,sym}(\omega)=-\frac{e^2\omega_{\rm c}^2}{4\omega}\nu
	\nonumber\\
	& \left(
		\Delta^{\rm sym}(\hbar\omega+\hbar\omega_{\rm c})+\Delta^{\rm sym}(\hbar\omega-\hbar\omega_{\rm c})
	\right).
	\label{eq:symsigma}
\end{align} 
Here, $\Delta^{\rm sym}(x)$ 
is a positive function that satisfies
\begin{align}
	&\Delta^{\rm sym}(x)=\Delta^{\rm sym}(-x) \mbox{ for } \forall x, 
	\label{eq:symcondition2}\\
	&\int_{-\infty}^{\infty} dx \Delta^{\rm sym}(x)=1, 
	\label{eq:symcondition1}\\
 	&\int_{|x| \geq y}dx\Delta^{\rm sym}(x)\ll 1 \ {\rm for}\ \forall y \gg \Gamma.
	\label{eq:symcondition4}
\end{align}
Substituting Eq.\ (\ref{eq:symsigma}) into Eq.\ (\ref{eq:deltasigmainter}), we find that the contribution of this symmetric broadening to $\Delta\sigma_{xy}^{\rm inter}$ is $\mathcal{O}(\epsilon^2\times e^2/h)$. 

Next, we consider the case where $\Im\sigma_{xy}^{\rm inter}(\omega)$ is broadened asymmetrically as
\begin{align}
	&\Im\sigma_{xy}^{\rm inter,asym}(\omega)=-\frac{e^2\omega_{\rm c}^2}{4\omega}\nu
	\nonumber\\
	& \left(
		\Delta^{\rm asym}(\hbar\omega+\hbar\omega_{\rm c})+\Delta^{\rm asym}(\hbar\omega-\hbar\omega_{\rm c})
	\right). 
	\label{eq:asymsigma}
\end{align} 
Here, $\Delta^{\rm asym}(x)$
is a positive function that satisfies
\begin{align}
	&\Delta^{\rm asym}(x)\neq\Delta^{\rm asym}(-x) \mbox{ for } \exists x, 
	\label{eq:asymcondition2}\\
	&\int_{-\infty}^{\infty} dx \Delta^{\rm asym}(x)=1, 
	\label{eq:asymcondition1}\\
 	&\int_{|x| \geq y}dx\Delta^{\rm asym}(x)\ll 1 \ {\rm for}\ \forall y \gg \Gamma.
	\label{eq:asymcondition4}
\end{align}
Substituting Eq.\ (\ref{eq:asymsigma}) into Eq.\ (\ref{eq:deltasigmainter}), we find that the contribution of this asymmetric broadening to $\Delta\sigma_{xy}^{\rm inter}$ is $\mathcal{O}(\epsilon\times e^2/h)$. 

From these results, we find that contributions of broadenings to $\Delta\sigma_{xy}^{\rm inter}$ are at most $\mathcal{O}(\epsilon \times e^2/h)$. 

\subsection{Loss of the spectrum}
The other type of deviation is the loss of the spectrum. 
In considering this issue, it is helpful to use the moment sum rule: 
\begin{align}
	\int_{-\infty}^{\infty}\frac{d\omega}{\pi}\Re\sigma_{\rm L(R)}(\omega)=\frac{ne^2}{m}=\frac{e^2}{h}\omega_{\rm c}\nu. 
	\label{eq:sumrule}
\end{align}
Here, 
\begin{align}
	\Re\sigma_{\rm L(R)}(\omega)
:=\Re\sigma_{xx}(\omega)\pm\Im\sigma_{xy}(\omega)
\label{eq:LR}
\end{align}
is proportional to the optical absorption spectra for the left (L) and right (R) circularly polarized light. 
As has been done in Eqs.~(\ref{eq:sigma_decmposed1}) and (\ref{eq:sigma_decmposed2}), we decompose it as 
\begin{align}
	\Re\sigma_{\rm L(R)}^{\rm intra}(\omega)&=\Re\sigma_{\rm L(R)}(\omega)\theta(\omega_{\rm c}/2-|\omega|)
	\label{eq:sigmaLR_decmposed1}
	\\
	\Re\sigma_{\rm L(R)}^{\rm inter}(\omega)&=\Re\sigma_{\rm L(R)}(\omega)\theta(|\omega|-\omega_{\rm c}/2))
	\label{eq:sigmaLR_decmposed}
\end{align}
using the Heaviside step function, $\theta(x)$.

Notably, $\Re\sigma_{xx}^{\rm intra}(\omega)$ is an even function, and is nonzero only at around $\omega=0$ of width $\mathcal{O}(\Gamma/\hbar)$\cite{SCBA}.  
Since only the states in the Landau level in which the chemical potential lies are relevant to the intra-band transition, the typical value of $\Re\sigma_{xx}^{\rm intra}(\omega)$ is independent of $\hbar\omega_{\rm c}$ and hence of $\epsilon$. 
From these and the fact that $\Im\sigma_{xy}^{\rm intra}(\omega)$ is an odd function part of $\Re\sigma_{\rm L(R)}^{\rm intra}(\omega)$, we obtain
\begin{align}
	\int_{-\infty}^{\infty}\frac{d\omega}{\pi}\Re\sigma_{\rm L(R)}^{\rm intra}(\omega)
	&=
	\int_{-\infty}^{\infty}\frac{d\omega}{\pi}\Re\sigma_{xx}^{\rm intra}(\omega)
	\nonumber\\
	&\simeq
	\mathcal{O}\left(\epsilon^0\times \frac{e^2}{h}\right)
	\times
	\mathcal{O}\left(\frac{\Gamma}{\hbar}\right) > 0. 
	\label{eq:sumruleintra}
\end{align}
From this and Eqs.~(\ref{eq:sumrule}), (\ref{eq:sigmaLR_decmposed1}) and (\ref{eq:sigmaLR_decmposed}) we obtain 
\begin{align}
	\int_{-\infty}^{\infty}\frac{d\omega}{\pi}\Re\sigma_{\rm L(R)}^{\rm inter}(\omega)
    &=\frac{e^2}{h}\omega_{\rm c}\nu\left(1-\mathcal{O}(\epsilon)\right),
	\label{eq:sumA}
\end{align}
which should be compared with the pure case, Eq.\ (\ref{eq:sumrule}).
The oscillator strength of the inter-band transitions decreases by $\mathcal{O}(\epsilon)$ due to the appearance of the intra-band transitions. 
This effect makes $\Im\sigma_{xy}^{\rm inter}(\omega)$ deviate from $\Im\sigma_{xy}^{\rm (pure)}(\omega)$ by $\mathcal{O}(\epsilon \times e^2/h)$.
Therefore, from Eq.~(\ref{eq:deltasigmainter}), the contribution of the loss of the spectrum to $\Delta\sigma_{xy}$ is $O(\epsilon \times e^2/h)$.

\subsection{Estimation of $\Delta\sigma_{xy}^{\rm inter}$}

We have seen that $\sigma_{xy}$ deviates from $\sigma_{xy}^{\rm (pure)}$ by the broadening and the loss of the spectrum.
The deviation by the former is at most $\mathcal{O}(\epsilon \times e^2/h)$, while that by the latter is $\mathcal{O}(\epsilon \times e^2/h)$.
Therefore, by taking account of both effects, we obtain Eq.\ (\ref{eq:deltasigmainter_order}). 

\section{Order of magnitude of $\Delta\sigma_{xy}^{\rm intra}$}
\label{ap:intra}
In the plateau region of the integer quantum Hall effect, the value of $\sigma_{xy}$ is determined only by the physical constant and is independent of $\epsilon$. 
Therefore, 
\begin{align}
	\Delta\sigma_{xy}=\mathcal{O}\left(\epsilon^0\times \frac{e^2}{h}\right). 
\end{align}
From this evaluation and Eq.\ (\ref{eq:deltasigmainter_order}), we obtain Eq.\ (\ref{eq:deltasigmaintra_order}) for 
 $\Delta\sigma^{\rm intra}_{xy} =\Delta\sigma_{xy}-\Delta\sigma^{\rm inter}_{xy}$.

\end{document}